\documentclass[a4paper]{jpconf}
\usepackage{graphicx}
\usepackage{amsfonts,amsmath,amssymb,bm,graphicx,cite,makeidx,multicol,color}
\usepackage{slashed}
\begin{document}
\title{Electromagnetic neutrinos: New constraints and new effects in oscillations}

\author{Alexander Studenikin$^{1, 2}$}

\address{$^1$ Department of Theoretical Physics, Moscow State University, 119992 Moscow, Russia}
\address{$^2$ Dzhelepov Laboratory of Nuclear Problems, Joint Institute for Nuclear Research, 141980 Dubna, Russia}

\ead{studenik@srd.sinp.msu.ru}
\begin{abstract}
 A short overview of neutrino electromagnetic properties
 with focus on existed experimental constraints and future prospects is presented.
  The related new effect in neutrino flavour and spin-flavour oscillations in the transversal
  matter currents is introduced.
\end{abstract}

{$\bf{1.\ Neutrino \ magnetic \ and \ electric \ dipole \ moments.}$}
The most well understood and studied among the neutrino electromagnetic properties \cite{Giunti:2014ixa,Studenikin:2008bd, Studenikin:2018vnp} are the dipole magnetic and electric moments.
In a minimal extension of the Standard Model  the diagonal magnetic  moment of a Dirac neutrino is given \cite{Fujikawa:1980yx} by
$\mu^{D}_{ii}
  = \frac{3e G_F m_{i}}{8\sqrt {2} \pi ^2}\approx 3.2\times 10^{-19}
  \Big(\frac{m_i}{1 \ \mathrm{eV} }\Big) \mu_{B},
  $
$\mu_B$ is the Bohr magneton. The Majorana neutrinos in the mass basis can have  only transition
(off-diagonal) magnetic
moments  $\mu^{M}_{i\neq j}$. However, in the flavour basis
the diagonal magnetic and electric moments of the
Majorana neutrinos can be nonzero.

The most stringent constraints on the effective neutrino
magnetic moment are obtained with the reactor antineutrinos
(GEMMA Collaboration \cite{GEMMA:2012})
$\mu_{\nu} < 2.9 \times 10^{-11} \mu_{B}$,
and solar neutrinos (Borexino Collaboration \cite{Borexino:2017fbd})
${\mu}_{\nu_e}\leq 2.8 \times
10^{-11} \mu _B.$
It should be noted, that
the magnetic
and electric moments measured in these experiments are not those of
massive neutrinos, but they are effective moments and they account
for the neutrino mixing and  oscillations
during the propagation between source and detector \cite{Grimus:1997aa, Beacom:1999wx}.
For the recent and detailed study of the neutrino electromagnetic characteristics
dependence on neutrino mixing see \cite{Kouzakov:2017hbc}.

An astrophysical bound (for both
Dirac and Majorana neutrinos) is provided
\cite{Raffelt-Clusters:90, Viaux-clusterM5:2013, Arceo-Diaz-clust-omega:2015}
by observations of the properties of globular cluster stars:
$\Big( \sum _{i,j}\left| \mu_{ij}\right| ^2\Big) ^{1/2}\leq (2.2{-}2.6) \times
10^{-12} \mu _B.$
A general and termed model-independent upper bound on the Dirac neutrino
magnetic moment, that can be generated by an effective theory beyond
a minimal extension of the Standard Model, has been derived in
\cite{Bell:2005kz}: $\mu_{\nu}\leq
10^{-14}\mu_B$. The corresponding limit for transition moments of Majorana neutrinos is much weaker \cite{Bell:2006wi}.


\ \ \ \ \ \ In the theoretical framework with $CP$ violation a neutrino
can have nonzero electric moments $\epsilon_{ij}$. In the laboratory neutrino
scattering experiments for searching $\mu_{\nu}$ (for instance, in the GEMMA experiment)
the electric moment $\epsilon_{ij}$ contributions interfere with
those due to $\mu_{ij}$. Thus, these kind of experiments also provide constraints
on $\epsilon_{ij}$. The astrophysical bounds on $\mu_{ij}$
are also applicable for constraining $\epsilon_{ij}$ (see \cite{Raffelt-Clusters:90, Viaux-clusterM5:2013, Arceo-Diaz-clust-omega:2015} and \cite{Raffelt:2000kp}).

{\it }
{$\bf{2.\ Neutrino \ electric \ millicharge.}$}
There are extensions of the Standard Model that allow for nonzero
neutrino electric millicharges. This option can be provided by
not excluded experimentally possibilities for hypercharhge dequantization or
another {\it new physics} related with an additional $U(1)$ symmetry
peculiar for extended theoretical frameworks (for the detailed discussion and corresponding
references see \cite{Giunti:2014ixa}). Neutrino millicharges
are strongly constrained on the level $q_{\nu}\sim 10^{-21} e_0$
($e_0$ is the value of an electron charge) from neutrality of the hydrogen atom.

 A nonzero neutrino millicharge $q_{\nu}$ would contribute to the neutrino electron scattering in the terrestrial experiments. Therefore, it is possible to get bounds on $q_{\nu}$ in the reactor antineutrino
 experiments. The most stringent reactor antineutrino constraint
 $q_{\nu}\leq 1.5 \times 10^{-12} e_0$
 is obtained in \cite{Studenikin:2013my} (see also \cite{PDG2016})
 with use of the GEMMA experimental data \cite{GEMMA:2012}.

A neutrino millicharge might have specific phenomenological consequences
in astrophysics because of new electromagnetic processes are opened
due to a nonzero charge (see \cite{Giunti:2014ixa,Raffelt:1996wa}). Following this line, the most stringent astrophysical constraint on neutrino millicharges
$q_{\nu}\leq 1.3 \times 10^{-19} e_0$
 was obtained in \cite{Studenikin:2012vi}. This bound
follows from the impact of the {\it neutrino star turning} mechanism ($ST\nu$) \cite{Studenikin:2012vi} that can be considered  as a {\it new phenomenon} end up with a pulsar rotation frequency
shift engendered by the motion of escaping from the
star neutrinos along curved trajectories due to millicharge interaction with a constant
magnetic field of the star.

{\it }
{$\bf{3.\ Neutrino \ charge \ radius \ and \ anapole \ moment.}$}
Even if a neutrino millicharge is vanishing, the electric form factor
$f^{ij}_{Q}(q^{2})$ can still contain nontrivial information about
neutrino electromagnetic properties. The corresponding electromagnetic characteristics is
determined by the derivative of $f^{ij}_{Q}(q^{2})$ over $q^{2}$  at
$q^{2}=0$ and is termed neutrino charge radius,
$\langle{r}_{ij}^{2}\rangle
=-
6
\frac{df^{ij}_{Q}(q^{2})}{dq^{2}} \
_{\mid _ {q^{2}=0}}
$ (this is indeed the charge radius squared, see \cite{Giunti:2014ixa} for the detailed discussions).
Note that for a massless neutrino the neutrino charge radius is the only
electromagnetic characteristic that can have nonzero value. In the Standard Model
the neutrino charge radius and the anapole moment are not defined separately,
and there is a relation between these two values: $a = - \frac{\langle{r}^{2}\rangle}{6}$.

A neutrino charge radius contributes to the neutrino scattering cross section on electrons and thus
can be constrained by the corresponding laboratory experiments \cite{Bernabeu:2004jr}.
In all papers, published before our study \cite{Kouzakov:2017hbc}, it was claimed
 that the effect of the neutrino
charge radius can be included just as a shift of the vector coupling constant $g_V$
in the weak
contribution to the cross section.
However, as it has been recently demonstrated in \cite{Kouzakov:2017hbc} within the direct calculations of
the elastic neutrino-electron scattering cross section accounting for all possible neutrino electromagnetic characteristics
and neutrino mixing, this is not the fact. The neutrino charge radius dependence of the cross section
is more complicated and there are, in particular, the dependence on the interference terms of the type
$g_{V}\langle{r}_{ij}^{2}\rangle$ and also on the neutrino mixing.

{\it }
{$\bf{4.\ Future \ prospects .}$}
 The foreseen progress in constraining neutrino electromagnetic characteristics is related, first of all, with the expected new results from the GEMMA experiment measurements of the reactor antineutrino cross section on electrons at the Kalinin Power Plant. A new set of data is expected to arrive next year. The electron energy threshold will be as low as $350 \ eV$ ( or even lower, up to $\sim 200 \ eV$). This will provide possibility to test the neutrino magnetic moment on the level of $\mu_\nu \sim 0.9 \times 10^{-12} \mu_B$ and also to test the millicharge on the level of $q_{\nu} \sim 1.8 \times 10^{-13} e_0$ \cite{Studenikin:2013my}.

The current constraints on the flavour neutrino charge radius $\langle{r}_{e,\mu,\tau}^{2}\rangle\leq 10^{-32} - 10^{-31} \ cm ^2$
from the scattering experiments differ only by 1 to 2
orders of magnitude from the values $\langle{r}_{e,\mu,\tau}^{2}\rangle\leq 10^{-33} \ cm ^2$ calculated within the minimally extended Standard Model with right-handed neutrinos
\cite{Bernabeu:2004jr}. This indicates that the minimally extended Standard Model neutrino charge radii could be experimentally tested in the near future.

Note that there is a need to re-estimate experimental constraints on
$\langle{r}_{e,\mu,\tau}^{2}\rangle$  from the scattering experiments following
new derivation of the cross section \cite{Kouzakov:2017hbc} that properly accounts for the interference of the weak and charge radius electromagnetic interactions and also for the neutrino mixing.

Recently constraints on  charged radii  have been obtained
\cite{Caddedu:2018prd} from the analysis of the data on coherent
elastic neutrino-nucleus scattering obtained in the COHERENT experiment
\cite{Akimov:2017ade}. In addition to the customary diagonal
charge radii $\langle{r}_{e,\mu,\tau}^{2}\rangle$,
also the neutrino transition (off-diagonal) charge radii have been constrained
in \cite{Caddedu:2018prd} for the first time:
$
\left(|\langle r_{\nu_{e\mu}}^2\rangle|,|\langle r_{\nu_{e\tau}}^2\rangle|,|\langle r_{\nu_{\mu\tau}}^2\rangle|\right)
< (22,38,27)\times10^{-32}~{\rm cm}^2$. These constraints have been included to
the recent update of the Review of Particle Properties \cite{PDG2016}.

Quite recently the potential of current and next generation of coherent elastic
neutrino-nucleus scattering experiments in probing neutrino electromagnetic
interactions has been also explored \cite{Miranda:2019wdy}.


For the future progress in studying (or constraining) neutrino electromagnetic properties
a rather promising claim was made in  \cite{deGouvea:2012hg}. It was shown that
even tiny values of the Majorana neutrino transition moments
would probably be tested in future high-precision experiments with the astrophysical neutrinos.  In particular,
observations of supernova fluxes  in the JUNO  experiment (see
\cite{An:2015jdp,Giunti:2015gga,Lu:2016ipr})
may reveal the effect of  collective  spin-flavour oscillations  due to the Majorana neutrino transition moment $\mu^{M}_\nu \sim 10^{-21}\mu_B$. There are indeed other new possibilities for neutrino
magnetic moment visualization in extreme astrophysical environments
considered recently \cite{Grigoriev:2017wff,Kurashvili:2017zab}.

In the most recent paper \cite{Cadeddu:2019qmv} we have proposed an experimental setup to observe coherent elastic neutrino-atom scattering using electron antineutrinos from tritium decay and a liquid helium target. In this scattering process with the whole atom, that has not beeen observed so far, the electrons tend to screen the weak charge of the nucleus as seen by the electron antineutrino probe.
 Finally, we study the sensitivity of this apparatus to a possible electron
 neutrino magnetic moment and we find that it is possible
 to set an upper limit of about
$\mu_{\nu} < 7 \times 10^{-13} \mu_{B},$
at 90 \%  C.L.,  that is more than one order of magnitude smaller than
the current experimental limits from GEMMA \cite{GEMMA:2012} and Borexino
\cite{Borexino:2017fbd}.

{$\bf{5. \ New \ phenomenon \ in \ neutrino \ oscillations \ in \ transversal
\ matter \ current.}$}
In the presence of a
magnetic field the neutrino flavour oscillations pattern is modified. The presence of
a magnetic field can engender neutrino spin and also spin-flavour oscillations.
A review on this issue can be found in \cite{Giunti:2014ixa} (see also \cite{Popov:2019nkr}).
As it has been shown in \cite{Popov:2019nkr} in the presence of a magnetic field  it is not possible to consider the neutrino flavour and spin oscillations
  as separate phenomena. On the contrary, there is an inherent communication between two.
  In particular, the amplitude of the neutrino flavour oscillations is modulated by the
  magnetic frequency $\omega_{B}=\mu B_{\perp}$.

It was shown in \cite{Studenikin:2004bu} that neutrino spin oscillations
can be induced not only by the neutrino interaction with a  magnetic field, as it was believed
before, but also by neutrino interactions with matter in the case when there
is a transversal matter current or matter polarization. A detailed study of the effect
is given in \cite{Pustoshny:2018jxb}. The main result of the discussions in \cite{Studenikin:2004bu,Pustoshny:2018jxb}
  is the conclusion on the equal role that the transversal magnetic field ${\bm B}_{\perp}$ and the
  transversal matter current $\bm{j}_{\perp}$ plays in generation of the neutrino spin and spin-flavour oscillations.

  From these observations, and also taking into account the mentioned above
inherent communication between flavour and spin oscillations \cite{Popov:2019nkr},
  \textbf{ we predict a new phenomenon of the modification of
  the flavour neutrino oscillations probability in moving matter under the condition of non-vanishing
  matter transversal current $\bm{j}_{\perp}=n \bm {v}_{\perp}$ }.
Given the similarity of the action of the magnetic field
$B_{\perp}$ and transversal matter current  $\bm{j}_{\perp}$
the flavour neutrino oscillation probability accounting for the effect of moving
matter  can be expressed as follows:
$P^{(j_{||}+j_{\perp})}_{\nu_{e}^L \rightarrow \nu_{\mu}^L} (t) =  \left(1 - P_{\nu_{e}^L \rightarrow \nu_{e}^R}^{(j_{\perp})} - P_{\nu_{e}^L \rightarrow \nu_{\mu}^R}^{(j_{\perp})}\right)
P_{\nu_{e}^L \rightarrow \nu_{\mu}^L}^{(j_{||})}$,
where
$P_{\nu_{e}^L \rightarrow \nu_{\mu}^L}^{(j_{||})}(t) =
\sin^2 2\theta _{eff}\sin^2 \omega_{eff}t \ \ $ 
is the flavour oscillation probability in moving matter \cite{Grigoriev:2002zr},
 $\omega_{eff}=\frac{\Delta m^2 _{eff}}{4p_{0}^\nu}$,
$\theta _{eff}$
and $\Delta m^2 _{eff}$  are the corresponding quantities modified by
the presence of moving matter
(note that in the definition of $\theta _{eff}$
and $\Delta m^2 _{eff}$ only the longitudinal component of matter motion matters).
Following an analogy with the studies performed in \cite{Studenikin:2004bu,Pustoshny:2018jxb},
we derive the probability of the neutrino spin and spin-flavour oscillations  engendered by
the transversal current $\bm {j}_{\perp}$:
\begin{equation}
P^{j_{\perp}}_{\nu_{e}^L \rightarrow \nu_{k}^R}(t)=
\frac{\Big(\frac{\eta}{\gamma}\Big)_{ek}^2 {v}^{2}_{\perp}}
{\Big(\frac{\eta}{\gamma}\Big)_{ek}^2 {v}^{2}_{\perp}+
\Big(\frac{\Delta M}{{\widetilde{G}}n} (1-\delta_{ek})- (1-{\bm v}
{\bm \beta})\Big)^2}
\sin^2\omega^{j_{\perp}}_{ek}t, \ k=e,\mu, \ \delta_{ee}=1, \ \delta_{e\mu}=0,
\end{equation}
for the notations used see \cite{Pustoshny:2018jxb}.
The discussed new effect of the modification of the
flavour oscillations $\nu_{e}^L\Leftarrow (j_{||}, j_{\perp})\Rightarrow \nu_{\mu}^L$ probability
is the result of an interplay of oscillations on a customary
flavour oscillation frequency in moving matter $\omega_{eff}$ and
two additional oscillations with changing the neutrino polarization (the neutrino spin $\nu_{e}^L\Leftarrow (j_{\perp}) \Rightarrow \nu_{e}^R$ and spin-flavour
$\nu_{e}^L\Leftarrow (j_{\perp}) \Rightarrow \nu_{\mu}^R$ oscilations) that are
governed by two characteristic frequencies :
$\omega^{j_{\perp}}_{ek}={\widetilde{G}}n
  {\sqrt{\Big(\frac{\eta}{\gamma}\Big)_{ek}^2 {v}^{2}_{\perp}
  +
  \Big(\frac{\Delta M}{{\widetilde{G}}n}(1-\delta_{ek})-
 (1-{\bm v}{\bm \beta})\Big)^2}}$, $k=e,\mu$.
The  interplay of neutrino oscillations on the introduced different frequencies
can have important consequences for neutrino fluxes  in astrophysical environments.

\end{document}